**Locally-Induced Stark Shifts of Collective Excitonic Modes in Polyradical Aggregates**

*Amandeep Sagwal*, Rodrigo Cezar de Campos Ferreira, Petr Kahan, Maximilian Rödel, Jindřich Nejedlý, Jiří Doležal, Martin Švec**

A. Sagwal, R. C. de Campos Ferreira, P. Kahan, J. Doležal, M. Švec

Institute of Physics, Czech Academy of Sciences; Cukrovarnická 10/112, Praha 6 CZ16200, Czech Republic

A. Sagwal

Faculty of Mathematics and Physics, Charles University; Ke Karlovu 3, CZ12116 Praha 2, Czech Republic

R. C. de Campos Ferreira, P. Kahan, J. Doležal, M. Rödel, J.Nejedlý, M. Švec

Institute of Organic Chemistry and Biochemistry, Czech Academy of Sciences; Flemingovo náměstí 542/2. Praha 6 CZ16000, Czech Republic

P. Kahan

Faculty of Nuclear Sciences and Physical Engineering, Czech Technical University in Prague, Břehová 7, 115 19 Prague, Czech Republic

*E-mail: (sagwal@fzu.cz; svec@fzu.cz)

**Abstract**

Active control of dark long-lived excitonic states in molecular aggregates using local electric fields is a pivotal challenge for advancing nanoscale optoelectronics and quantum device engineering. This experimental study investigates the collective excitonic states in aggregates composed of radical chromophores. With the strong optical enhancement provided by tip-enhanced photoluminescence (TEPL) spectroscopy, bright and dark excitonic modes are observed emerging due to interexciton coupling and induce changes in their spectra with the electric field locally applied within the nanocavity gap. Proportionally scaling Stark shifts are revealed as well as the emission peak sharpening of the dark states and a divergent behavior of the bright states in asymmetric measurement positions of the nanocavity above the aggregates. The observed complex behavior is discussed in terms of influence of the field, molecule arrangement, nanocavity coupling, dark mode lifetimes and electrostatic charge inhomogeneities in the clusters. This sensitivity to the external parameters demonstrates an effective means of control over radical excitonic aggregates.

## 1 Introduction

Aggregates of chromophores with prevailing Coulombic coupling are known to host well-defined collective excitonic modes in contrast to systems with an admixture of charge-transfer interactions.**[1-5]** In such systems, superradiant modes may emerge - providing the advantage of enhanced radiative rates; on the other side the dark states, with their characteristically long lifetimes, hold the promise to support efficient energy transfer among chromophores, which can be of high importance for the development of future optoelectronic devices and may help to better understand fundamental energy harvesting mechanisms in living systems.**[6-8]** In contrast to the neutral chromophore aggregates that have been studied quite extensively in this context recently, reports on systems composed of radical anions are scarce, **[9-12]** despite their potential to host more exotic excitonic properties. Since individual anion radicals often exhibit higher polarizability, they are prone to intermolecular static charge redistribution and screening, which can render them responsive to external fields, useful for excitonic switching.**[13]** Their open-shell character may also bring spin-dependent functionality, which can lead to higher fluorescence efficiency due to different relaxation pathways **[14]** and magnetic field-activity due to the unpaired electron.**[15-17]** To tackle the problem of creating and investigating such stable radical anion aggregates, one of the promising strategies is to precisely assemble well-defined clusters of precursors with strong acceptor or donor character on insulating crystalline surfaces.

Exploring the collective excitonic modes of aggregates requires specialized approaches that combine high spatial and spectral resolution at cryogenic conditions, because of the fine (Davydov) splitting and energy ordering of spectral fingerprints of the excitonic modes of individual clusters that vary in geometry.**[18-21]** This is beyond reach of optical spectroscopy working in the far-field, hindered by the Abbe diffraction criterion, inhomogeneous broadening, validity of Kasha's rule and weak optical activity of dark modes due to the symmetry-imposed selection rules or forbidden electronic transitions.**[5,22]** In addition, achieving local control over the excitons requires a highly defined geometry and environment of the nanostructures. Emerging near-field spectroscopic techniques employ an extremely localized electromagnetic field confined in the junction of a scanning probe microscope to make the otherwise forbidden transitions optically accessible.**[23-27]** With this approach, demonstrations of the sensitivity to the dark states were performed on various neutral phthalocyanine assemblies on thin crystalline insulating layers with outstanding precision. It permitted investigation of their coherence and entanglement using scanning tunneling microscope-induced electroluminescence (STM-EL).**[28-31]** In our previous work, we have applied STM-EL to investigate small aggregates of widely-studied electron acceptors, perylene dicarboxylic anhydride (PTCDA) which can be stabilized in dominantly radical state in the nanocavity, and revealed hints that switching could be induced with local application of an electric field.**[32]** However, the STM-EL method requires the application of bias above a threshold to drive the excitations by injection of charges, and consequently it cannot track the excitonic response to a continuously-tuned field in the nanocavity and evaluate Stark and vibronic shifts. This can be overcome by the TEPL technique that allows accessing the excitonic modes without the need of an external bias.**[33-34]** Notably, this approach remains currently underexploited.

In this work, we investigate the effect of the electric field on the photon-response of dominantly anionic PTCDA aggregates confined in STM plasmonic nanocavity at submolecular level, focusing on the characteristic behavior of the bright and dark exciton modes. We confirm the controllability of the collective modes by the electric field, by the aggregate geometry and the nanocavity positioning in various clusters. We discuss the complex behavior of the bias-dependent TEPL fingerprints in the context of a simple model that allows us to estimate the Stark shifts of the individual chromophores, their mutual Coulombic coupling, and the influence of the nanocavity.

## 2 Results and Discussion

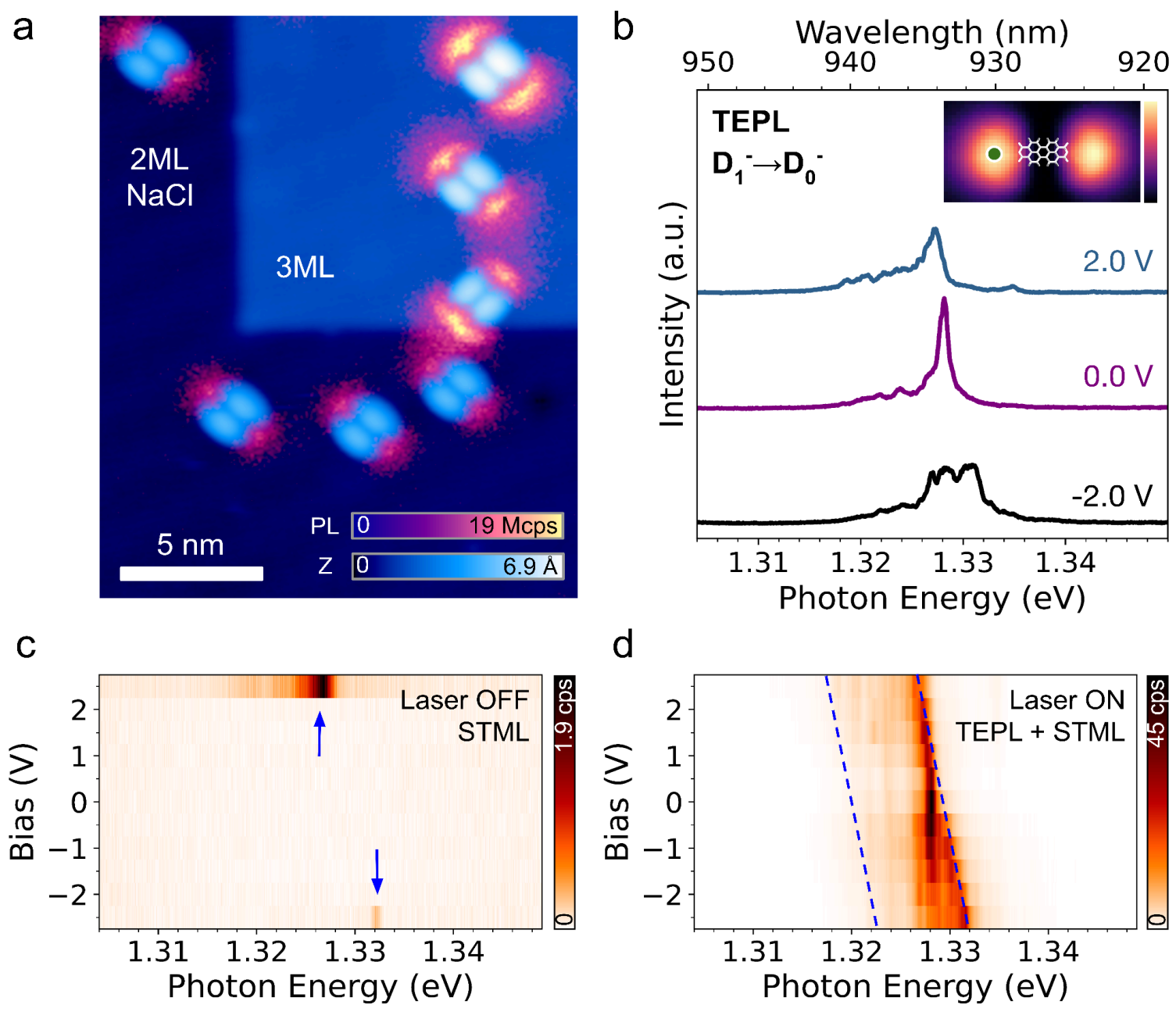

**Figure 1.** (a) STM constant-current topography and tip-enhanced luminescence of PTCDA on 2-3 ML NaCl/Ag(111) crystal surface. Scanning conditions: 900 mV, 3 pA, 16.9 x 23.8 nm$^2$. The excitation wavelength of the laser was 785 nm (1.579 eV) with 1.1 mW total output power. The photon rate map was taken simultaneously using APD with a 925/25nm filter and superimposed over the topographic image using a different color scale. (b) Representative TEPL spectra acquired at -2.0, 0.0 and 2.0 V. The inset shows a constant-height TEPL intensity map for the $D_1^- \rightarrow D_0^-$ radiative transition, shown for the 1.328 eV energy at zero bias. The green dot marks the location of the point-spectra measurements in (c) and (d). The size of the map is 4.8 x 2.9 nm$^2$, 48 x 29 points. (c, d) Spectral intensity as a function of bias without and with laser excitation, respectively. The vertical arrows in (c) denote the onsets of the electroluminescence. The blue dashed line in (d) marks a common trend and high-energy cutoff of the bias-dependent TEPL spectra.

First, we studied bias-dependent spectral fingerprints of an individual PTCDA molecule. A composite image of surface topography (Z) and tip-enhanced photoluminescence signal (TEPL) of individual PTCDA molecules adsorbed on 2 and 3 ML of NaCl on Ag(111) is shown in Figure 1a. Above each of the molecules, a characteristic two-lobe luminescence pattern is observed. This pattern originates from the molecule's radiative electronic transition from its first excited state to the ground state ($D_1^- \rightarrow D_0^-$), which is coupled to the nanocavity.**[14]** At 3 ML the luminescence yield is stronger compared to 2 ML; it can be attributed to a better decoupling from the metal substrate and a more efficient radiative decay rate.**[35-38]** The point-spectra of the $D_1^- \rightarrow D_0^-$ emission in Figure 1b taken above the carbonyl-terminations, are presented for three different bias values at -2.0, 0.0 and 2.0 V. Such strong variations among these spectra indicate an effect of the electric field in the tip-sample junction. With respect to the spectrum obtained at zero bias, which has one narrow dominant peak with small sidebands, the -2.0 and 2.0 V spectra show a general blue- and redshift of few meV, respectively and the spectral envelope is notably widened for the spectrum at -2.0 V.

Detailed bias-dependence measurements with and without illumination allow us to put the bias-induced Stark shift and complex restructuring of the spectral envelope into the context of the previously reported STM electroluminescence measurements.**[32,39-40]** The 2D heatmaps created from individual spectra in Figures 1c,d provide a comparison of the two scenarios on the same system. At -2.5 V the electroluminescence (in Figure 1c) manifests an onset at -2.5 V, producing a very sharp peak at 1.332 eV. At the positive limit of 2.5 V, a broader electroluminescence peak is found at ≈5.4 meV higher energy. No measurable signal is detected between the onsets, in this setup. On the other hand, TEPL (Figure 1d) provides spectra for all biases in the range and reveals that the overall spectral envelope (delimited by the dashed lines) shifts constantly by approximately -0.9 meV/V. The electroluminescence and TEPL overlap at the ±2.5 V with their most intense features, however TEPL has a stronger overall yield and significantly broader envelope. We attribute this broadening to the wider range of vibrational modes that are excited and coupled to the electronic transition under optical excitation, compared to excitation mediated by charge carrier injection. In addition, a closer look at the spectra in the entire bias range reveals transient sharp spectral components within the broad envelope that are barely changing the energy with the applied bias. The most intense is a prominent peak occurring at 1.328 eV, in the range -1.0 to 1.0 V. We attribute these peaks tentatively to the effects of resonant Raman scattering that may match vibronic transitions - as it was suggested in previous study of single phthalocyanines.**[41]** Such resonant behavior, however, does not compromise the TEPL ability to detect the Stark shifts, particularly of the coupled dark excitonic states, which tend to be very sharp in energy, as it will be demonstrated in the following investigation of selected aggregates.

Next, to investigate the photophysical properties of PTCDA anion aggregates, we created the simplest model systems, i.e. dimers. We achieved that with molecular manipulation and thermal diffusion strategies.**[32]** Firstly, we manipulated two molecules to be on the same $Cl^-$ row along the ⟨110⟩ direction on the NaCl(001) surface, with their longer axes aligned collinearly. The measured intermolecular distance of 1.95 nm corresponds to 5 $Cl^-$ lattice spacings, as shown in Figure 2a. The zero-bias TEPL spectrum measured at the peripheral end of either molecule of the dimer has a peak at 1.328 eV, nearly identical in shape as a single molecule. However, the TEPL intensity spatial distribution at this energy over the dimer (inset of Figure 2b) does not correspond to a simple superposition of the individual molecular emission patterns. Instead, the strongest intensity is localized at the outer sides of the molecules and the inner region between them exhibits a considerably weaker contribution. Furthermore, the spectrum obtained in the central region, in addition to the 1.328 eV peak, has a sharp feature emerging at 1.331 eV. The TEPL map at this energy shows higher intensity between the two molecules. This is an expected behavior in two identical coulombically coupled excitons with a dominant dipolar character, forming a prototypical J-aggregate.**[42]** Therefore we can attribute the two bands at 1.328 and 1.331 eV to the emission from the bonding (symmetric, superradiant) and antibonding (antisymmetric, dark) modes of the coupled $D_0^- \leftrightarrow D_1^-$ excitons of the molecules, respectively, which we denote as $|\widetilde{0}\rangle$ and $|\widetilde{1}\rangle$ according to their order in energy.

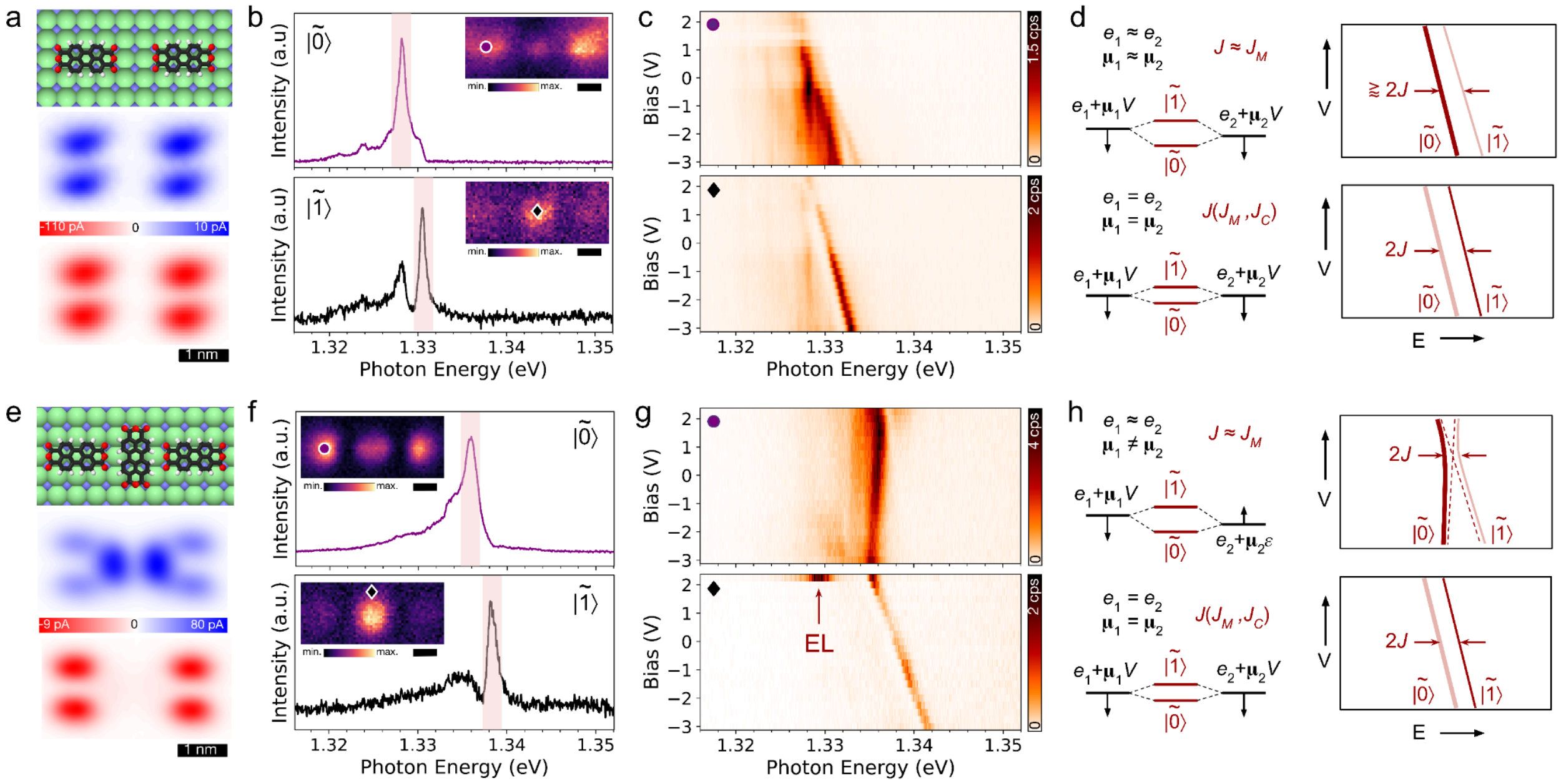

**Figure 2.** (a) Model of PTCDA dimer on NaCl and constant-height current image taken at 1.0 and -0.8 V, size 3.8 x 2.0 $nm^2$. ($Cl^-$ and $Na^+$ atoms are marked green and purple, respectively). (b) TEPL point-spectra of the dimer at 0 V, acquired at the peripheral and central sites (marked by purple dot and black diamond, respectively). The insets show TEPL intensity maps for the 1.328 eV and 1.331 eV, respectively, measured at constant height at 0 V, with size 6.0 x 2.6 $nm^2$, 60 x 26 points (c) Bias-dependent TEPL spectral intensity represented as heatmaps for the sites marked in (b). (d) Schematic diagram for the on-site energies and eigenenergy levels for rationalization of the bias-induced shifts of the collective excitonic modes in the PTCDA dimer. (e) Model of PTCDA trimer aggregate and constant height images at ±1 V, with size 3.8 x 2.0 $nm^2$. (f) TEPL point-spectra acquired at the periphery and center region of the trimer at 0 V. The insets show TEPL signal maps for 1.336 eV and 1.338 eV, respectively, measured at constant height at 0 V, with size 6.0 x 2.6 $nm^2$, 60 x 26 points. (g) Bias-dependent TEPL spectral intensity heatmap at the sites marked in (f). The arrow denotes the electroluminescence onset of the middle molecule at 2.25 V. (h) Schematic diagrams for the on-site energies and the collective mode levels for the PTCDA trimer used for the explanation of the divergent shift of the bonding mode ($|\tilde{0}\rangle$) in response to the nanocavity positioning above the trimer.

The bias-dependence of the $|\widetilde{0}\rangle$ emission measured at the dimer periphery (Figure 2c) (purple circle) closely resembles the single-molecule, including the energy shift and restructuring. In contrast, the lineshape of $|\widetilde{1}\rangle$ taken in the center of the dimer (black diamond) is considerably sharper in the entire bias range, with the peak energy position following a linear trend (approximately -0.8 meV/V). Similarly as in Figure 2b where the spectrum of the $|\widetilde{1}\rangle$ contains a hint of the $|\widetilde{0}\rangle$ state, the bias-dependent spectra show a mixture of the two states, too. This is a result of a nonzero nanocavity coupling with the $|\widetilde{0}\rangle$ state, regardless of the measurement position above the aggregate. When the distance of the parallel molecules in the dimer is varied, again by manipulation or a using thermal self-assembly (see Figure S1 and Figure S2), the near-correspondence of the $|\widetilde{0}\rangle$ and $|\widetilde{1}\rangle$ Stark shifts is preserved; only the energy splitting narrows as the distance is increased, due to a diminishing coupling among the two excited states. The linear response of both excitonic modes to the applied electric field can be rationalized as a result of simultaneous shifting of the on-site energies of both chromophores, as illustrated in Figure 2d and discussed further below.

In a similar system (assembled by thermal diffusion), where a third PTCDA is wedged between the two chromophores in a perpendicular orientation and stabilized by the carboxylic-hydrogen interactions (see Figure 2e), we observe a specific change of the apparent Stark shift for the bright mode measured at the periphery. The zero bias TEPL spectra in Figure 2f at the peripheral and the central sites of such a symmetrical trimer show two peaks at 1.336 and 1.338 eV, respectively. In this symmetrical trimer, the current maps and d$I$/d$V$ (Figure S3c) prove that the central PTCDA unit maintains an equilibrium neutral charge within a broad bias range up to 2.0 V, in accord with our previous findings.**[32]** Considering that only the peripheral anions couple and contribute to the collective states, resolved in the TEPL intensity distribution maps (insets of Figure 2f) we can attribute the peaks again to the $|\widetilde{0}\rangle$ and $|\widetilde{1}\rangle$ states. However, the bias-dependence of the $|\widetilde{0}\rangle$ spectra measured at the periphery is different from the case of the dimer; the lineshape is sharper and more uniform across the biases, with a reversed Stark shift of approximately 0.2 meV / V and a trend change above 1.5 V. The $|\tilde{1}\rangle$ emission is very similar to the case of dimers, sharp and shifting with a similar magnitude (-1.2 meV / V). At biases above 2.0 V, where a reversible switch of the central neutral molecule to an anion is expected, we observe its electroluminescence appearing at about 1.329 eV (denoted by a red arrow in Figure 2g).

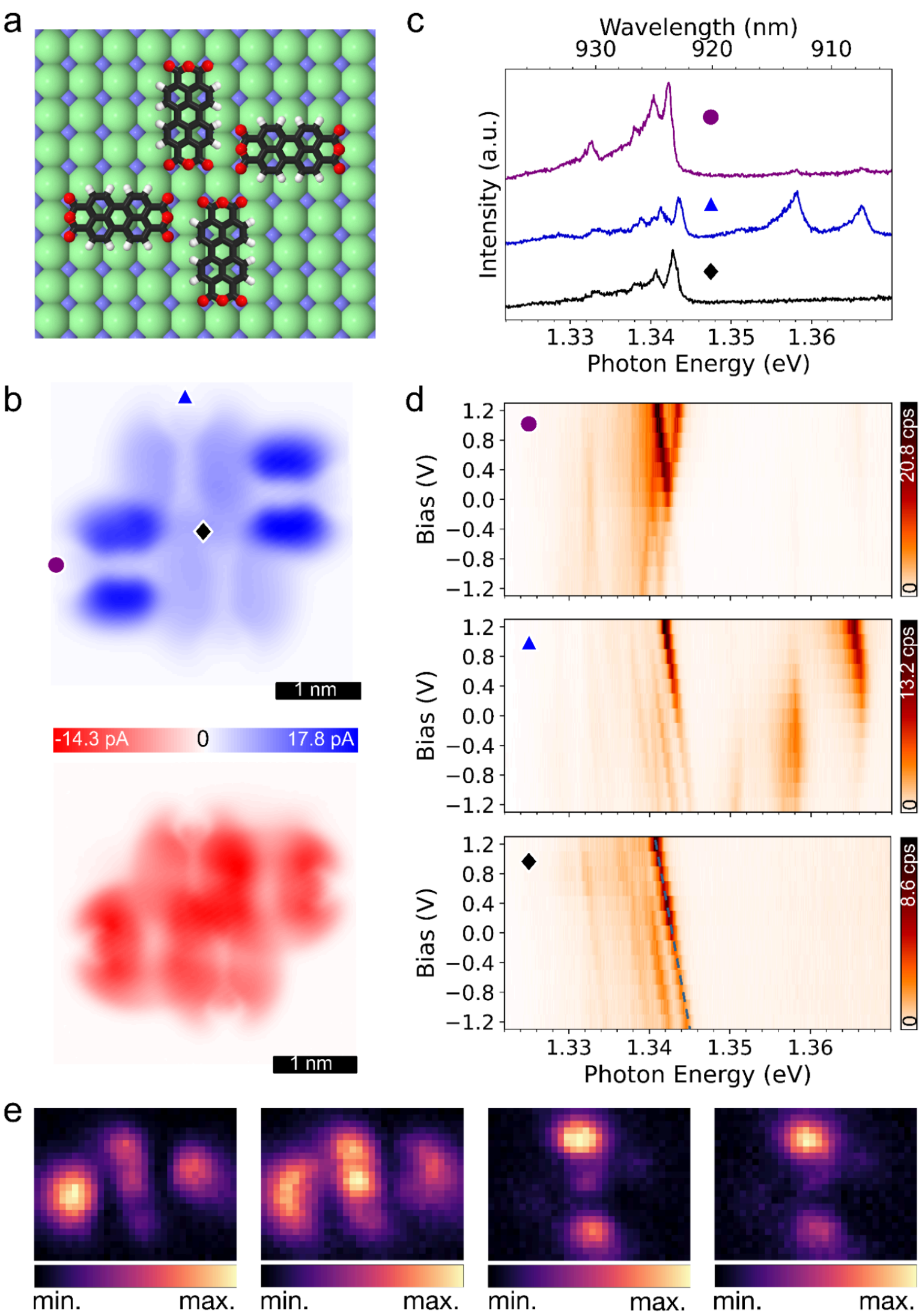

**Figure 3.** (a) Model of PTCDA tetramer on NaCl (Cl- and Na+ atoms are marked green and purple, respectively). (b) constant-height current image taken at +0.5V and -0.25 V, 5pA, size 3.56 x 3.56 $nm^2$. (c) TEPL point-spectra of the dimer at 0 V, acquired at the two peripheral and the central positions (denoted by purple dot, blue triangle and black diamond, respectively, also in (b)). (d) Bias-dependent TEPL spectral intensity represented as heatmaps for the sites marked in (b). (e) TEPL intensity maps for the 1.340, 1.342, 1.356 and 1.365 eV, measured at constant height at 0 V, with size 6.0 x 2.6 $nm^2$, 60 x 26 points.

We extend the investigation to larger and more complex aggregates, here a particular tetramer with a twofold rotationally-symmetrical ($C_2$) chiral structure (Figure 3a). It comprises two pairs of equivalent molecules due to the symmetry of the arrangement. The current maps and dI/dV spectra (Figure 3b and Figure S3d, respectively) show that the electronic signatures of the two types of molecules in the cluster are strongly shifted in energy compared to monomers and the peripheral molecules in the trimers, but their general characteristics corresponds to PTCDA anions, including the small broad feature between the positive and negative ion resonances of the dI/dV that was previously attributed to an increased dynamic conductance of a monomer under irradiation.**[14]** The electronic signature of the pair of equivalent molecules closer to the center of the cluster slightly differs from the other pair. It is worth mentioning that the current, as well as the dI/dV signal, are nanocavity position-dependent but carry a time-averaged information; therefore, various charge configurations must also be considered in the interpretation of the TEPL spectra, apart from the measurement site dependence. The TEPL spectra at zero bias in Figure 3c measured at three different representative locations above the tetramer (shown in Figure 3b) display multiple peaks that evolve in analogy to the dimers and trimers (Figure 3d). At the center point of the tetramer (black diamond), the bias-dependence shows a dominating sharp spectral feature with a pronounced negative Stark shift (-1.56 meV/V), accompanied by higher-energy parallel sidebands. When measured over one of the equivalent peripheral molecules farther from the center (denoted with purple circle), these sidebands seem to switch the polarity of the Stark shift (1.45 meV/V), resulting in an apparent crossover with the dominant band. Such overlap of the energies for the bonding and antibonding features is observed near zero bias, and is consistent with the photon map (Figure 3e) that shows a hybrid pattern of intensity localized at the periphery and in the aggregate center at the same time.

Based on these characteristics, the feature dominating most of the spectra of the tetramer can be attributed to the prominent excitonic antibonding mode of the system, while the sidebands at lower energies to the bonding-type coupling among all the molecules within the clusters along the longer dimension of the tetramer. Finally, when measuring at the periphery, in the direction of the shorter dimension (marked with blue triangles), broad peaks appear at higher energies above 1.35 eV, which we therefore assign to a dominant coupling among the two geometrically equivalent molecules along the shorter dimension of the tetramer, again based on the spatial correlation with the respective photon map in Figure 3e. Hallmarks of a similar behavior, including the change of the Stark shifts of spectral features when measured at the center and periphery of the aggregate, are also observed in a more complex aggregates (shown in Figure S4), although the spectral manifestations of the collective modes are less clear, most likely due to the higher amount of PTCDA units involved, coupling strength constrained by the geometry and the limited resolution of the method.

The antibonding collective modes of all the described structures manifest a fairly consistent proportional bias-induced shift between -1.2 and -1.6 meV/V regardless of the nanocavity position over the structures. The divergent and sharpened bonding modes of the trimer, tetramer and hexamer were observed with the nanocavity positioned above their respective peripheral parts, whereas no such effect was present on the simple dimers. The specific aspect of the dimers, compared to the other studied aggregates, is the absence of the interstitial PTCDA units between the peripheral molecules. For an elementary phenomenological rationalization of the observed peak shifts in the dimer in Figure 2c, we consider two electronic transitions with intrinsic on-site energies $e_1$ and $e_2$, coupled through the effective overall coupling $J$ (which is a nontrivial product of the interexciton coupling $J_M$ and an additional coupling mediated by the nanocavity $J_C$) and shifted by the applied bias $V$ with proportional factors $\mu_1$ and $\mu_2$, expressed as $E_{1,2} = e_{1,2} + V\mu_{1,2}$. The resulting eigenenergies will be simply given by **[6,43- 44]**:

$$E_{\pm} = \frac{E_1+E_2}{2} \pm \sqrt{\left(\frac{E_1-E_2}{2}\right)^2 + J^2}. \quad (1)$$

For the situation with nanocavity symmetrically placed above the center of the dimer, the intrinsic on-site energies should be equal, i.e. $e_1 = e_2$. The observed parallel shifting of the $|\tilde{0}\rangle$ and $|\tilde{1}\rangle$ modes with bias can be then understood as a result of a uniform polarization of the two sites regardless of the nanocavity position (for simplicity $\mu_1 = \mu_2$). Moving the nanocavity above one of the sites results in a small changes among the respective constant Lamb and Stark shifts (i.e. $e_1 \approx e_2$ and $\mu_1 \approx \mu_2$) and a reduced effective coupling $J$ due to the diminishing role of the nanocavity-mediated part $J_C$. The consequence would be a slightly smaller separation of the $|\tilde{0}\rangle$ and $|\tilde{1}\rangle$ modes and a small deviation from their parallel evolution, as shown in the diagram in Figure 2d. On the other hand, to use the same model to reconcile the diverging shift of the bonding mode of the trimer with nanocavity positioned at the periphery (Figure 2g), the bias-dependent on-site energies of the peripheral chromophores need to respond to the field differently. Specifically, one of the coefficients $\mu_1$ and $\mu_2$ must be strongly altered to reproduce the observed bias-dependence of the bonding state eigenenergy within the bias window (see Figure 2h). This scenario would result in an avoided-crossing at the point of the two on-site energy equality $E_1 = E_2$. Despite a lack of signal for the antibonding state in Figure 2g, a hint of the avoided crossing onset is visible around 1.8 V. We have attempted to optimize the parameters of this simple model to match the observed dependencies of the mode energies. The results are plotted with a good match over the data in Figure S5 and the corresponding parameters are included in Table S1. Since the divergent Stark shifts are observed for the bonding modes at the peripheries in the trimer, tetramer and hexamer (in contrast to the simpler dimers), we propose that this behavior primarily originates from a specific electrostatic screening effect at the perpendicularly oriented molecules between the coupled PTCDA units, regardless of their charge state.

On the basis of this simple model we can also get an elementary understanding of the behavior of the collective modes in the tetramer cluster. The dominating features in the symmetrical measurement position are the antibonding and bonding states of the two peripheral molecules.

Their parallel bias-dependence can be again understood as a result of their equally shifting on-site energies. Conversely, in the nanocavity position above one of the molecules, the crossover-like appearance can be seen as stemming from the diverging on-site energies. The broad spectral features above 1.35 eV associated with the pair of molecules closer to the center according to the photon maps indicate that the corresponding on-site energies are relatively strongly increased, most likely due to interactions with the surrounding PTCDA units. In this tetramer, it should be considered that the individual units could be switching between various charges (as the neutral and anion are close in energy), which could give rise to the observed additional multiplicity of the bands, however further measurement, analysis and simulations would be needed to corroborate this.

## 3 Conclusions

We have achieved with TEPL the direct and local electric-field tuning of the excitonic bonding- and antibonding-type eigenstates of coulombically-coupled PTCDA anion chromophore aggregates. A consistent, nearly linear Stark shifting of the delocalized states was found in several symmetrical clusters of various sizes in response to the electric field applied within the nanocavity gap between the tip and the sample. Measurements on the selected aggregates have revealed that the dark (antibonding-type) modes are generally sharper than their bright (bonding-type) counterparts. The nanocavity located above the high-symmetry points of the clusters produces very similar energy shifts for the dark and bright states, whereas measurements in off-symmetry peripheral positions lead to diverging shifts of the bright states. We attribute this phenomenon to a specific renormalization of the on-site Stark shift coefficients due to intraaggregate electrostatic screening. We can reconcile the observed behavior with a simple coupled-exciton model. Overall, our findings highlight the role of the nanocavity localization in the control of the collective excitonic modes (in particular the bright ones) and show the robustness of the dark state Stark shifts against it. This study also shows that engineered radical aggregates in plasmonic nanocavities can serve as templates for studying the electric field control of exciton coupling, exciton delocalization, and vibronic mixing with spatial and spectral control. This opens pathways toward the rational design of molecular-scale optoelectronic devices in which excitonic properties are tailored by nanoscale electric fields.

## 4 Experimental Section/Methods

Experiments were performed in an ultra high vacuum (UHV) environment using a low-temperature STM (Createc GmbH) operating at 6K. The Ag(111) single crystal was prepared by standard cycles of $Ar^+$ ion sputtering and thermal annealing. For decoupling layers, NaCl was thermally evaporated from a homebuilt evaporator at 607°C onto the half covered Ag sample maintained at 114°C for 4 minutes, resulting in formation of a partial mixture of 2-4 monolayers of NaCl islands. PTCDA molecules were deposited onto the NaCl/Ag(111) surface within the STM head for 3 minutes at 320°C. After deposition, the sample was left to thermalize for 1-2 minutes (estimated temperature 100-150K), in order to facilitate thermal diffusion and the formation of molecular aggregates. For dimer formation, the molecules were manipulated onto the same $Cl^-$ row of the NaCl surface, with their transition dipoles aligned collinearly. The tip was approached to the H-terminated region of the molecule opposite to the intended movement with 3 V and 20 pA feedback setpoint. The feedback loop was then turned off, the bias and setpoint were changed to 1.5 V and 5 pA. With these conditions, the feedback was turned back on and the area was rescanned to inspect the outcome. The procedure was repeated until the molecule was moved to the desired position. A plasmonic Ag tip was conditioned by applying voltage pulses and controlled indentations on a clean Ag(111) surface and tuned to cover both excitation and emission energy range. As the excitation source a single frequency continuous-wave (CW) infrared laser (785 nm) was used. The laser was focused into the tunneling junction by an aspheric lens through a confocal optical setup, employed also for the photoluminescence detection.**[14,35,45]** To acquire the photon rate a single photon avalanche photodiode from Perkin-Elmer (SPCM-AQR-15) was used. The TEPL spectra were acquired with Andor Kymera 328i spectrograph equipped with Newton 920 CCD detector. The differential conductance (dI/dV) spectra were measured using a standard lock-in approach with a modulation amplitude of 20 mV.

## Author Contributions

A.S. and M.S. conceived the experiment. J.N. purified the molecular precursors. R.C.C.F., A.S., P.K. and J.D. prepared the measurements, all authors have participated in the measurements.

A.S., R.C.C.F. and M.S. analyzed the data and all authors discussed the results and prepared the manuscript.

**Acknowledgements**

We are grateful for fruitful discussions with T. Neuman from the Institute of Physics. J.D. and M.R. acknowledge their funding through the IOCB postdoctoral fellowship program. The experimental measurements in this work were supported by the Czech Science Foundation Standard grant no. 22-18718S.

**Conflict of Interest**

The authors declare no conflicts of interest.

**Data Availability Statement**

The data that support the findings of this study are available from the corresponding author upon reasonable request.

https://doi.org/10.1021/acsnano.4c02105

**Supporting Information**

Additional supporting information can be found online in the Supporting Information section.

**TOC**

**Title:** Local electric-field and nanocavity control of excitonic modes in molecular aggregates.

Stark shifts in collective excitonic modes are investigated directly in radical chromophore aggregates using tip-enhanced photoluminescence using a scanning tunneling microscope junction. Varying electric fields are applied within the tip-sample nanocavity in distinct locations on different types of aggregates and the response of the excitonic modes is observed.

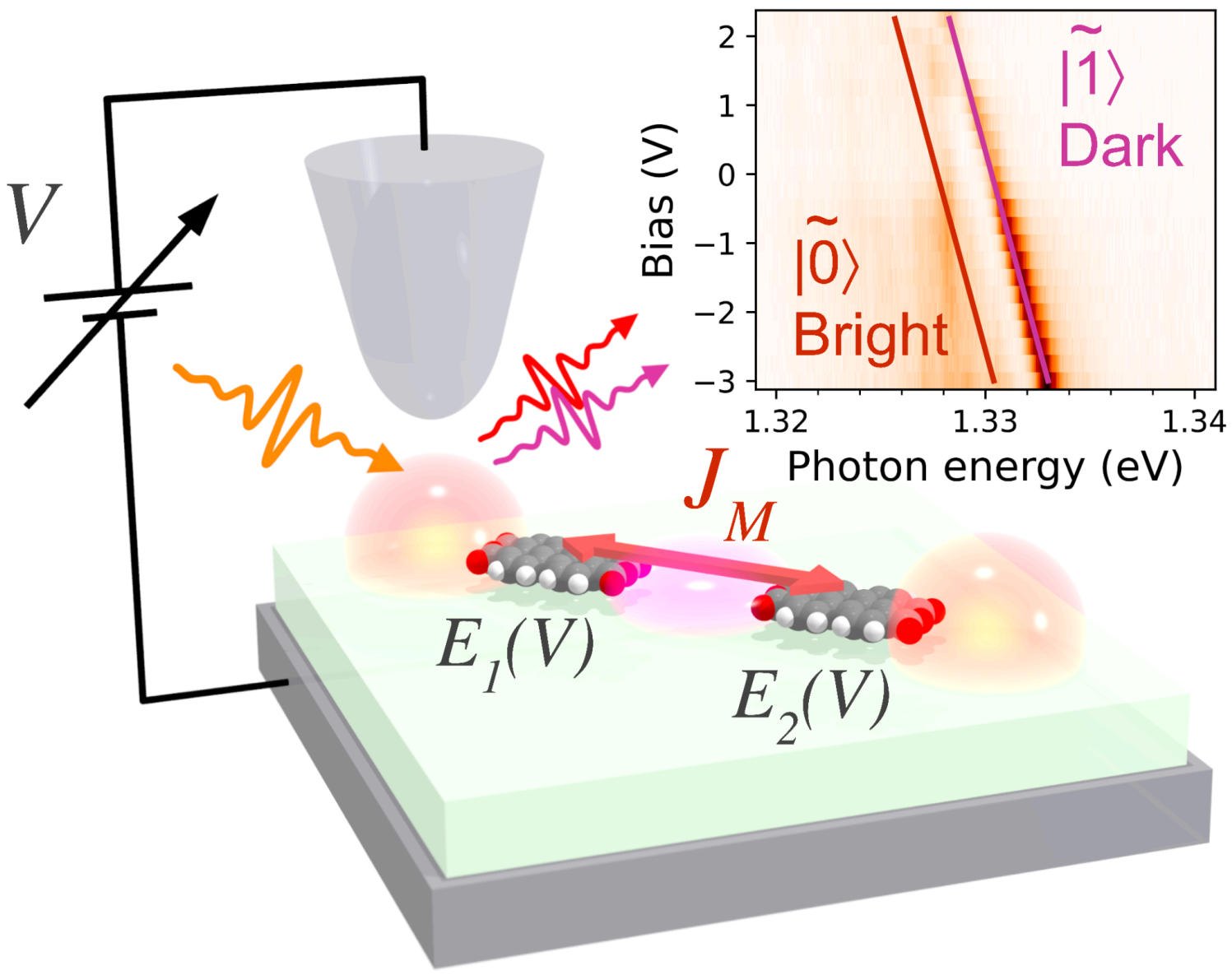

**Supplementary Information**

**Locally-Induced Stark Shifts of Collective Excitonic Modes in Polyradical Aggregates**

*Amandeep Sagwal*, Rodrigo Cezar de Campos Ferreira, Petr Kahan, Maximilian Rödel, Jindřich Nejedlý, Jiří Doležal, Martin Švec**

A. Sagwal, R. C. de Campos Ferreira, P. Kahan, J. Doležal, M. Švec

Institute of Physics, Czech Academy of Sciences; Cukrovarnická 10/112, Praha 6 CZ16200, Czech Republic

A. Sagwal

Faculty of Mathematics and Physics, Charles University; Ke Karlovu 3, CZ12116 Praha 2, Czech Republic

R. C. de Campos Ferreira, P. Kahan, J. Doležal, M. Rödel, J.Nejedlý, M. Švec

Institute of Organic Chemistry and Biochemistry, Czech Academy of Sciences; Flemingovo náměstí 542/2. Praha 6 CZ16000, Czech Republic

P. Kahan

Faculty of Nuclear Sciences and Physical Engineering, Czech Technical University in Prague, Břehová 7, 115 19 Prague, Czech Republic

*E-mail: (sagwal@fzu.cz; svec@fzu.cz)

**Figure S1: TEPL of parallel PTCDA dimer manipulated to various intermolecular distances.**

**Figure S2: A representative spontaneously-formed parallel PTCDA dimer with different orientation.**

**Figure S3: Differential conductance (dI/dV) of aggregates with illumination.**

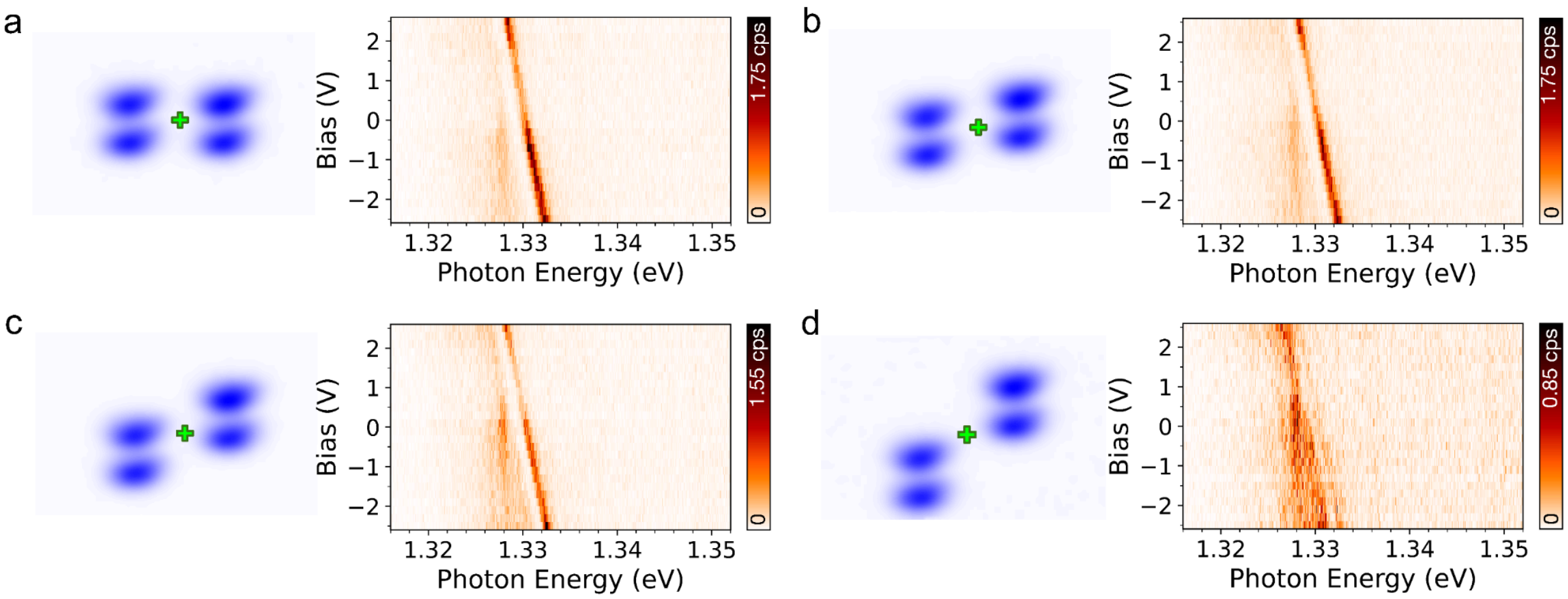

**Figure S1.** TEPL of parallel PTCDA dimer manipulated to various intermolecular distances. The distance between two parallel-oriented PTCDA was varied by moving one of them along the $\langle 110 \rangle$ direction on the NaCl(001) surface, resulting in four different geometries, corroborated by the constant-height current imaging at 1.2 V. The bias-dependent TEPL was measured in a center position (marked with green crosses) in order to simultaneously detect the $|\tilde{0}\rangle$ and $|\tilde{1}\rangle$ modes.

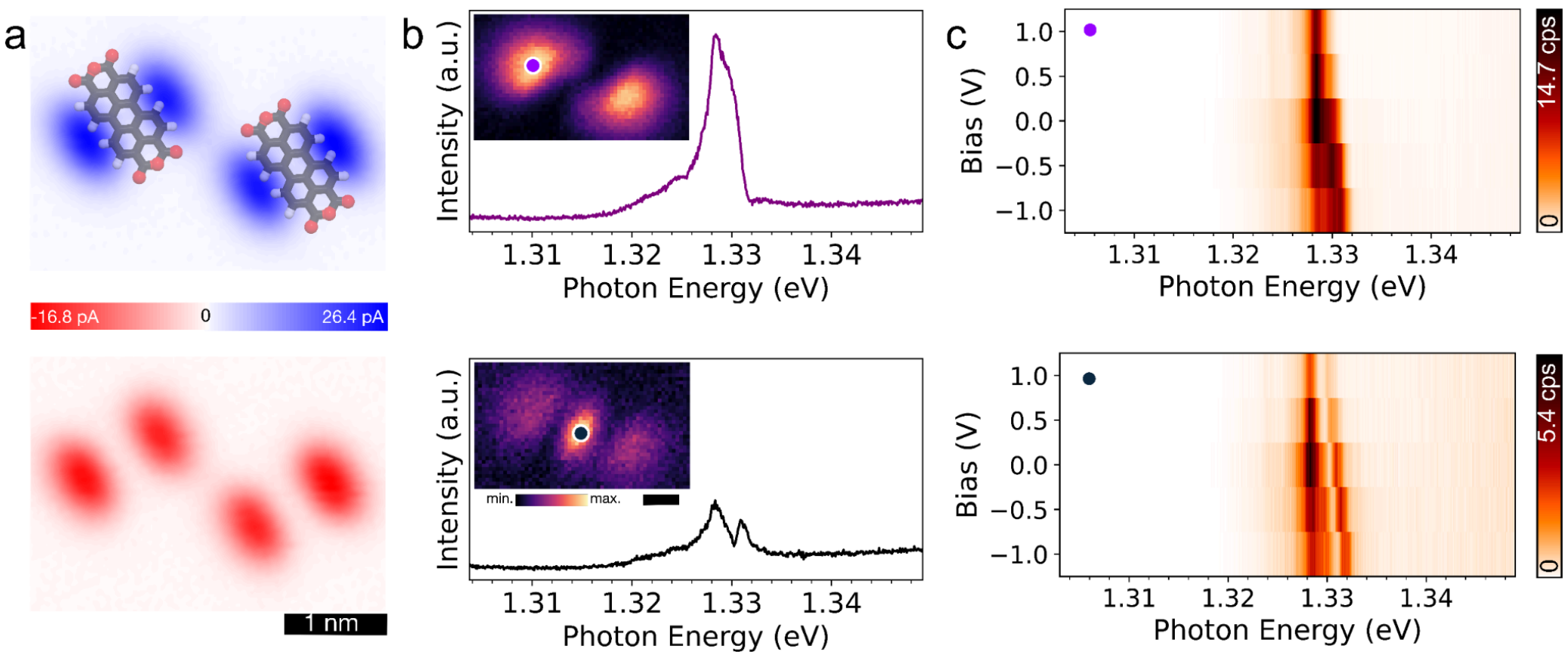

**Figure S2.** A representative spontaneously-formed parallel PTCDA dimer with different orientation. (a) Constant height scans of a dimer with ball-and-stick model overlay taken at +1.0

V and -0.7 V, 5 pA, size 3.5 x 2.5 nm$^2$. (b) TEPL spectra of both $|\tilde{0}\rangle$ and $|\tilde{1}\rangle$ modes taken at 0 V. The inset are the TEPL maps at 1.329 eV and 1.332 eV, respectively, acquired at 0.5 V, size 6.0 x 3.5 nm$^2$, 60 x 35 points. (c) Bias-dependent TEPL spectral intensity heatmap acquired at the peripheral and central position (marked with purple and black dots in inset of (b), respectively).

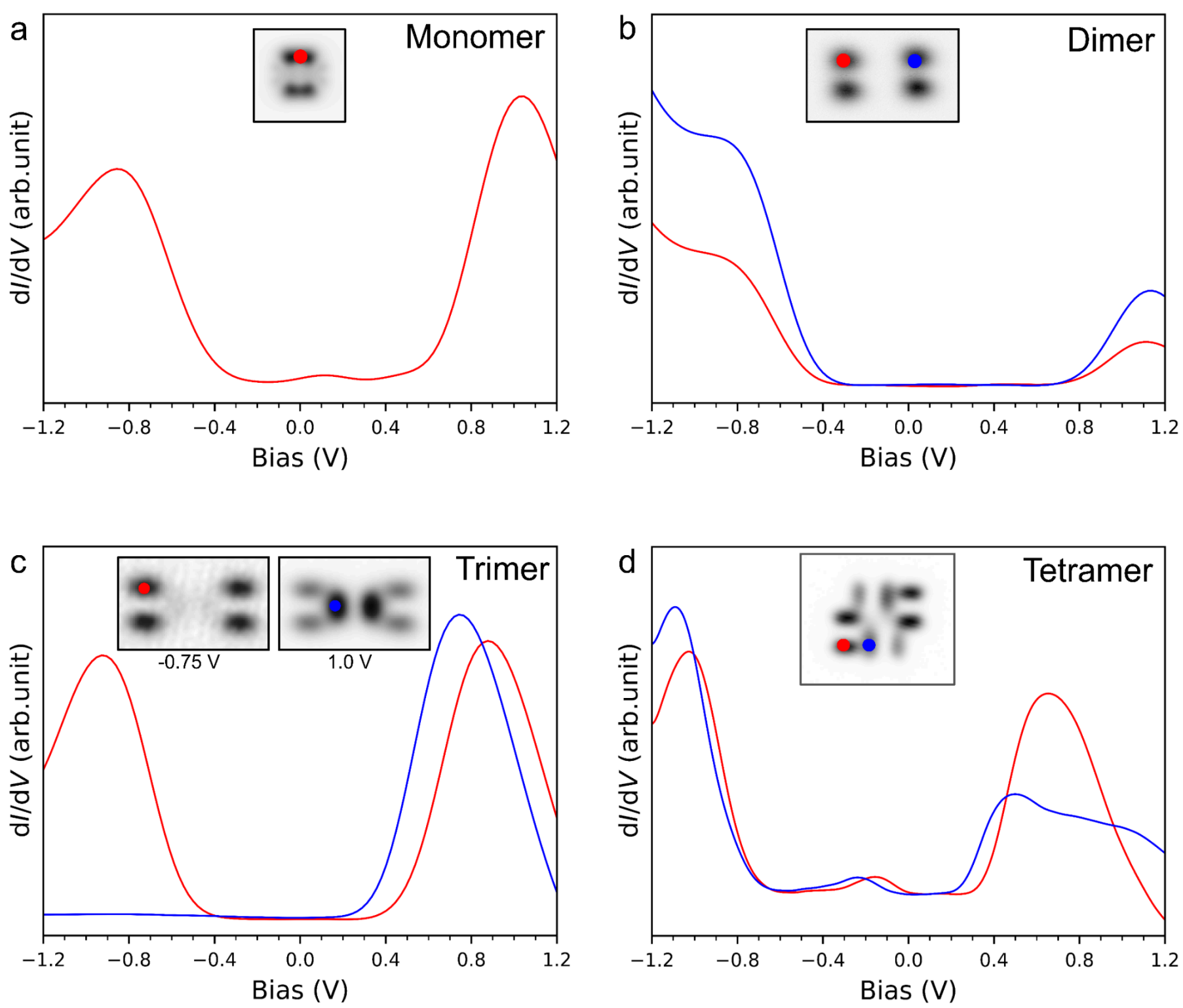

**Figure S3.** Differential conductance (dI/dV) of aggregates with illumination, taken at (a) monomer, (b) dimer, (c) trimer, and (d) tetramer. Constant-height current images are shown in the insets. The red and blue spectra were acquired at the positions marked by red and blue dots, respectively, in the corresponding inset. The dI/dV spectra were measured using the standard lock-in technique with bias modulation 20 mV and 927 Hz.

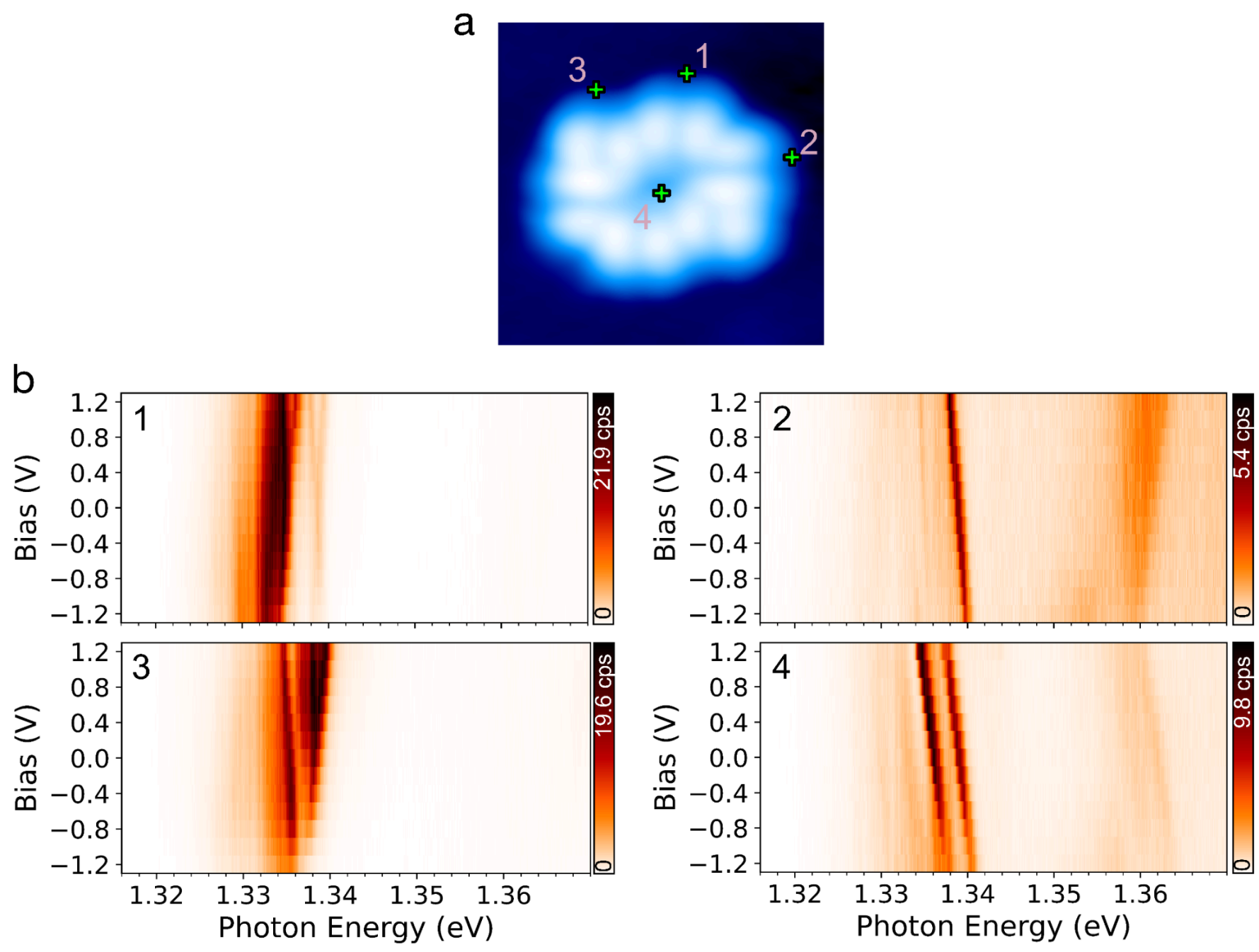

**Figure S4.** Hexamer PTCDA aggregates. (a) STM constant-current topography of a hexamer aggregate formed by thermal diffusion, with marked positions (1-4) of bias-dependent TEPL measurements. Scanning conditions: 1.2 V, 4 pA, size: 7 x 7 nm$^2$ (b) Bias-dependent TEPL heatmaps measured at positions marked in (a) at 0 V.

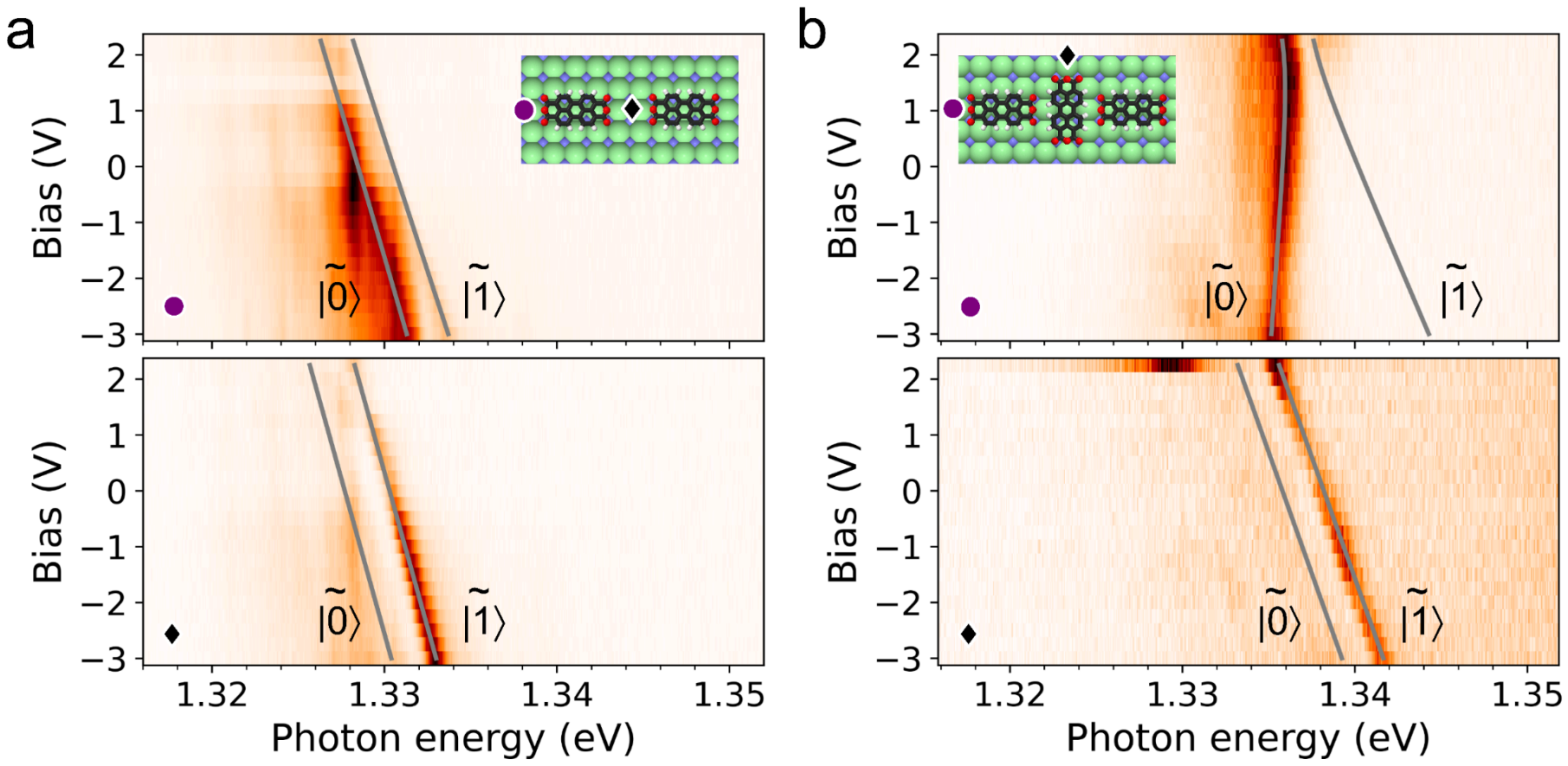

**Figure S5.** Matching the two-exciton model to bias-dependent TEPL for (a) the dimer and (b) the trimer. The trends of the bias-dependent peak energy positions for $|\tilde{0}\rangle$ (purple dot) and $|\tilde{1}\rangle$ (black diamond) modes were generated with the parameters of the model in Table S1. The insets show the ball-and-stick models of the aggregates.

| Aggregate | Site | $e_1$ (eV) | $e_2$ (eV) | $\mu_1$ (meV/V) | $\mu_2$ (meV/V) | $J$ (meV) |
|---|---|---|---|---|---|---|
| dimer | center | 1.329 | 1.329 | -0.90 | -0.90 | 1.30 |
| dimer | periphery | 1.330 | 1.329 | -1.10 | -0.90 | 0.90 |
| trimer | center | 1.337 | 1.337 | -1.20 | -1.20 | 1.20 |
| trimer | periphery | 1.340 | 1.336 | -1.40 | +0.25 | 0.90 |

**Table S1.** Parameters of the two-exciton model, used for the energy trends overlay in Figure S5. The positions (sites) are according to Figure 2.